\documentclass[12pt, fleqn]{article}
\usepackage[cp1251]{inputenc}
\usepackage{latexsym,amsfonts,amssymb}
\usepackage{graphicx}

\usepackage{amsbsy}
\usepackage{amsmath}
\usepackage{epsf}
\usepackage{cite}

\newtheorem{theo}{Theorem}

\newtheorem{remark}{Remark}
\newcommand{\bt}{\begin{theo}}
\newcommand{\et}{\end{theo}}
\newcommand{\bd}{\begin{displaymath}}
\newcommand{\ed}{\end{displaymath}}

\newcommand{\be} {\begin{equation}}
\newcommand{\ee} {\end{equation}}
\newcommand{\ba} {\begin{array}}
\newcommand{\ea} {\end{array}}
\newcommand{\bea}{\begin{eqnarray}}
\newcommand{\eea} {\end{eqnarray}}

\begin{document}

\begin{center}
 {\Large \bf A reaction-diffusion system with cross-diffusion:
 \\
 \vspace{0.3cm}
   Lie symmetry, exact solutions and their applications in \\ \vspace{0.3cm} the pandemic modeling}
\medskip

{\bf Roman Cherniha \footnote{\small  Corresponding author. E-mail: r.m.cherniha@gmail.com}}
  {\bf and  Vasyl' Davydovych \footnote{\small  E-mail:davydovych@imath.kiev.ua }}
 \\
{\it ~Institute of Mathematics,  National Academy
of Sciences  of Ukraine,\\
 3, Tereshchenkivs'ka Street, Kyiv 01004, Ukraine
}\\
 \end{center}

 \begin{abstract}
A nonlinear reaction-diffusion system with cross-diffusion  describing the COVID-19 outbreak   is
 studied  using the Lie symmetry method. A complete Lie symmetry classification is derived and  it is shown
that the system with correctly-specified parameters  admits  highly nontrivial Lie symmetry operators, which do not occur for all  known reaction-diffusion systems.  The symmetries
obtained are also applied for finding exact solutions  of the  system
in the most interesting case from applicability point of view. It is shown that the exact solutions derived possess all necessary properties for describing the pandemic spread under 1D approximation in space and lead to the distributions,
which qualitatively correspond to the measured data of the COVID-19 spread  in Ukraine.
\end{abstract}

\emph{Keywords:}
 reaction-diffusion system; cross-diffusion;  Lie symmetry;  exact solution; modeling  pandemic spread.

\section{ Introduction } \label{sec:1}

The outbreak of the  coronavirus called COVID-19 in China has attracted extensive attention of many  mathematicians working in mathematical modeling. The first papers were already published in February--April   2020 (see, e.g., \cite{ luo,china-19-02-20,shao,tian,efim-ushi, roda-michaelLi,ch-dav-preprint20}.
At the present time,  the COVID-19 outbreak is already  spread   over the world as a pandemic. There were  65.5~mln. coronavirus cases   and almost  1.5 mln.  deaths caused by this coronavirus up to date  December 2  \cite{meters}.


Nowadays, there are many mathematical models used to describe epidemic processes and
they can be found in any book devoted to mathematical models in biology
and medicine (see, e.g., \cite{brauer-12,k-r-2008,  diekman-hees-2000, mur2003}  and papers cited therein).
The paper \cite{kermack-1927} is one of the first papers in this direction. The authors created a model based on three ordinary differential equations (ODEs), which nowadays is called the SIR model. There are several  generalizations of the SIR model and the SEIR model (see the pioneering works \cite{anderson,dietz}), which involves four ODEs, is the most common among them. These two models are mostly used for numerical simulations in mathematical modeling the COVID-19 outbreak (see, e.g., \cite{china-19-02-20,tian,efim-ushi}).

On the other hand, one may note that the spread of many epidemic processes, including the COVID-19 pandemic, is often highly non-homogenous in space. This fact can be taken in different ways but the most common approach consists in  dividing  the large domain (say a country) into many small sub-domains (regions of the country) and  to apply the standard models based on ODEs to each sub-domain. However, there is another way -- to use the reaction diffusion equations in order to model the spread of the infected population as a diffusion process \cite{viguerie-2020, mammeri-2020} (see also earlier papers cited therein). A possible model was also  suggested in our previous work \cite{ch-dav-2020}. The model has the form
\begin{equation}\label{a1}\begin{array}{l} \medskip  u_t = d_1 \Delta u+u(a- bu^\gamma),  \\
v_t = d_2\Delta u +k(t)u,\end{array}\end{equation}
 where the lower subscript $t$ means differentiation with respect to (w.r.t.) this variable,
 $\Delta $ is
the Laplace operator,  $u=u(t,x,y)$  and $v=v(t,x,y)$ are two  unknown functions, $k(t)$ is the  given smooth positive function, $d_1$ and  $d_2$  are diffusivities.

The function $u(t,x,y)$ describes the density (rate) of the infected persons (the number of the  COVID-19 cases) in a vicinity of the point $(x,y)$,  while $v(t,x,y)$ means the density of the deaths from COVID-19. The diffusivity coefficients $d_1$ and  $d_2$ describe the random movement  of the infected persons, which lead to increasing the pandemic spread. Formally speaking, one may take $d_1=d_2$. However we believe  that $d_1>d_2$ because the movement of the infected persons leads firstly to higher rate of new COVID-19 cases but only some of them cause new deaths.
Each  coefficient in the reactions terms, $a, \ b, \ \gamma$  and $k(t)$, has the clear meaning described  and verified in \cite{ch-dav-2020} (see Pages 2 and 3 therein).

Of course, this model is an essential simplification because many factors causing the  spread of COVID-19 are not taking into account. In paper \cite{viguerie-2020}, for example, the authors construct the diffusion  model, which is essentially based on  the SEIR model. As a result, their model consists of five PDEs, which can be analyzed   only using numerical methods. Our idea was to construct a simpler model, which can be solved using analytical approaches, in particular, the Lie symmetry method \cite{ovs,ch-se-pl-book,bl-anco-10} and to show its applicability for the spread of the coronavirus pandemic. It is interesting to note that equations (3) and (5) from the model developed in \cite{viguerie-2020} under the natural assumptions  produce equations with a similar structure to those in (\ref{a1}). In fact,  the density  of infected population is proportional to that of the total living population  and is proportional to the exposed population density (generally speaking, the relevant coefficients are some  functions but we keep constants). Having such assumptions, one arrives at the system (in our notations)
\begin{equation}\label{a1**}\begin{array}{l} \medskip  u_t =  \nabla(d_1 u \nabla u)+u(a- bu),  \\
v_t = k(t)u,\end{array}\end{equation}
 The  difference between (\ref{a1}) with $\gamma=1 $  and (\ref{a1**}) consists  only in the diffusion terms. In our model, the diffusivity is taking to be a constant, while one is a linear function   in paper \cite{viguerie-2020}. Notably, the diffusivity  is a time-dependent function in paper \cite{mammeri-2020}.


The remainder of this  paper is organized as follows. In Section~\ref{sec:2}, a complete  Lie symmetry classification (LSC) (group classification)  of  system (\ref{a1}) is derived. In particular, we  have proved  that there are systems with correctly-specified parameters $d_1, \ d_2, \ a$ and $ \gamma$ when the system in question admits highly nontrivial Lie symmetry, which have no analogs for other known reaction-diffusion systems.  The results were obtained   using the Lie--Ovsiannikov method \cite{ovs},  which is a combination of the classical Lie method and the technique  for finding  equivalence transformations (ETs).  The modern description of this method, its extension and applications  can be found in \cite{ch-se-pl-book}~(Chapter~2).

In Section~\ref{sec:3}, exact solutions of the specified  system  of the form (\ref{a1}) are constructed using its Lie symmetry operators. In particular, the traveling wave type solution is derived and its applicability is extensively discussed. It is shown that this exact solution  describes adequately  the spread of the coronavirus pandemic provided  the 1D approximation of space is assumed.
 Finally, we  discuss the main  results of the paper  in the last section.

\section{ Main Results } \label{sec:2}

In this section, it is identified that the basic system (\ref{a1}) for the pandemic  modeling  possesses a very reach Lie symmetry depending on the parameters  $\gamma, \ d_1$ and  $d_2$  and the function $k(t)$.
First of all, we note that for the LSC we need only the restrictions   $d_1^2+d_2^2\neq0$ (otherwise the system in question degenerates into the ODE system, which was solved in \cite{ch-dav-2020}), $k\neq0$ (otherwise the system in question seems to be useless for applications) and  $b\neq0$ and $\gamma\neq0,-1$ (otherwise the system in question is linear, hence is also integrable).

First of all, we present a statement about the  group of ETs of system (\ref{a1}).
For this purpose we apply the technique, which was developed in \cite{akh-gaz-ibr91, ibr-tor-val91} (see also Section 2.3 in \cite{ch-se-pl-book}).

\begin{theo} \label{th1} The group of the continues ETs   transforming
  system (\ref{a1})  to that with the same structure, i.e.,
\begin{equation}\label{a*}\begin{array}{l} \medskip  \overline{u}_{\overline{t}} = \overline{d_1} \Delta \overline{u}+\overline{u}(\overline{a}- \overline{b}\overline{u}^{\overline{\gamma}}),  \\
\overline{v}_{\overline{t}} = \overline{d_2}\Delta \overline{u} +\overline{k}(\overline{t})\overline{u},\end{array}\end{equation} is the  infinite-parameter Lie group
generated by the transformations
 \begin{equation}\label{a3}\begin{array}{l} \overline{t} =\beta_1t+\alpha_0, \
\overline{x}=\beta_2\left(x\cos \beta_0 +y\sin \beta_0\right)+\alpha_1, \
\overline{y}=\beta_2\left(y\cos \beta_0 -x\sin \beta_0\right)+\alpha_2, \\
\overline{u}=\beta_3u, \ \overline{v}=\beta_4v+f(x,y), \\
 \overline{k}=\frac{\beta_4}{\beta_1\beta_3}\,k, \ \overline{d_1}=\frac{\beta^2_2}{\beta_1}\,d_1, \ \overline{d_2}=\frac{\beta^2_2\beta_4}{\beta_1\beta_3}\,d_2, \
 \overline{a}=\frac{a}{\beta_1}, \ \overline{b}=\frac{b}{\beta_1\beta_3^\gamma}, \ \overline{\gamma}=\gamma.
 \end{array}\end{equation}
 Here $\alpha_i \
(i=0,1,2)$ and $\beta_j \ (j=0,\dots,4)$    are the real group parameters with the restrictions
$\beta_1\beta_2\beta_4\neq0, \ \beta_3>0,$  and  $f(x,y)$ is an arbitrary  smooth function.
\end{theo}

\begin{remark} In order to obtain system (\ref{a*})  with the nonnegative parameters (this is the biologically motivated requirement explained above), the additional restrictions $\beta_1>0$ and $\beta_4>0$ should take place.
\end{remark}

\emph{Sketch of Proof of Theorem~\ref{th1}} is based on
 the known  technique for constructing the  group of ETs. It is nothing else but   a
modification of the classical Lie method. In the case of system (\ref{a1}), one should start from the  infinitesimal operator
\begin{equation}\label{2-1}\begin{array}{l}
E=\xi^0(t, x,u,v)\partial_t+\xi^1(t, x,u,v)\partial_x+\eta^1(t, x,u,v)\partial_u+\eta^2(t, x,u,v)\partial_v\\ +
\zeta(t,x,u,v,k)\partial_k+\mu^1\partial_{d_1}+\mu^2\partial_{d_2}+\mu^3\partial_{a}+
\mu^4\partial_{b}+\mu^5\partial_{\gamma} \end{array}
\end{equation}
being  $\xi^0, \ \xi^1, \ \eta^1,\ \eta^2 \ $ and $\zeta$ to-be-determined functions, while $\mu^i \ (i=1,\dots,5)$ to-be-determined constants. The operator $E$ involves the additional terms with the coefficients $\mu^i$  and $\zeta$, because $d_1, \ d_2, \ a, \ b, \ \gamma$ and $k(t)$ should be treated as  a new variables.

In order to find the operator $E$,  we should apply  Lie's invariance criteria to the system of equations consisting of
(\ref{a1}) and a set of differential consequences of  $k(t)$  w.r.t. the variables $t,\ x,\ u,\ v$.
 Of course, each consequence is equal to zero, excepting $\frac{\partial k}{\partial t}=k'(t)$ (the latter is not useful because it is identity). As a result, we obtain a multicomponent system  consisting equations from (\ref{a1})  and primitive equations like $\frac{\partial k}{\partial x}= 0$. Applying to this system Lie's invariance criteria, i.e., the second prolongation of the infinitesimal operator $E$, for deriving the system of determining equations, the coefficients  $\xi^0, \ \xi^1, \ \eta^1, \eta^2, \ \zeta$ and $\mu^i \ (i=1,\dots,5)$ were found. They have the form
 \begin{equation}\label{2-2}\begin{array}{l}
\xi^0=C_1t+C_2, \ \xi^1=C_3x+C_4y+C_5, \
\xi^2=C_3y-C_4x+C_6, \medskip \\
\eta^1=C_7u, \ \eta^2=C_8v+h(x,y),\ \zeta=\left(C_8-C_1-C_7\right)k, \
 \mu^1=\left(2C_3-C_1\right)d_1, \medskip \\
 \mu^2=\left(2C_3-C_1+C_8-C_7\right)d_2,\
\mu^3=-C_1a,\  \mu^4=-\left(C_1+C_7\gamma\right)b,\  \mu^5=0,
\end{array} \end{equation}
where $C_i, \ i=1,\dots,8$ are arbitrary constants and $h(x,y)$ is an arbitrary smooth function.
 The operator (\ref{2-1})  with the coefficients (\ref{2-2})  generates the Lie group (\ref{a3}).

 \emph{The sketch of the proof is now completed.}

 \medskip

 In order to provide a complete LSC of system (\ref{a1}), one should to identify the principal algebra of invariance (see definition, for example, in \cite{ch-se-pl-book}, page 23) from the very beginning. In fact,  system (\ref{a1}) involves an arbitrary function $k$ and several parameters (some of them can vanish). Thus, it should be considered as  a class of systems of partial differential equations (PDEs), if one is going to provide a rigorous LSC.

 \begin{theo} \label{th2*}
The  principal algebra of invariance of system (\ref{a1}) is infinite-dimensional  Lie algebra generated by the  operators:
\begin{equation}\label{a2} \partial_x, \ \partial_y, \ y\,\partial_x-x\,\partial_y,\ F(x,y)\,\partial_v,
\end{equation} where $F(x,y)$ is an arbitrary smooth function.
\end{theo}

 The proof of this statement can be derived in different two ways. The direct approach consists in application    of the classical Lie method to system (\ref{a1}), assuming that all parameters are arbitrary. The second way is useful
 if the group of ETs is known. So, having Theorem 1, we simply calculate when transformations (\ref{a3}) transform
 (\ref{a1}) in itself, i.e., system (\ref{a*}) coincides with (\ref{a1}). The result immediately leads to formulae (\ref{a2}).

 Now we present two main theorems, which completely solve the LSC problem for (\ref{a1}).
 It turns out  that there are two essentially different cases, $d_1\neq0$ and $d_1=0$, leading to absolutely different results.

\begin{theo} \label{th2}
System (\ref{a1}) with $d_1\neq0$ admits the extension of the  principal algebra (\ref{a2}) only in five cases.
These cases and the corresponding Lie symmetry operators are as follows

1) $k(t)=1$\,: $\partial_t$;

2) $a=0, \ k(t)=\frac{1}{t}$\,:    $2t\,\partial_t+x\,\partial_x+y\,\partial_y-\frac{2}{\gamma}\left(u\partial_u+v\partial_v\right)$;

3) $a=0, \ d_2=0, \ k(t)=t^p, \ p\neq-1,0$\,: $2t\partial_t+x\,\partial_x+y\,\partial_y-\frac{2}{\gamma}\left(u\partial_u+(1-\gamma p-\gamma)v\partial_v\right);$

4)  $d_2=0, \ k(t)=e^{pt}, \ p\neq0$\,:  $\partial_t+pv\partial_v$;

5)  $d_2=0, a=0, \ k(t)=1$\,:  $\partial_t,$ \ $2t\partial_t+x\,\partial_x+y\,\partial_y-\frac{2}{\gamma}\left(u\partial_u+(1-\gamma)v\partial_v\right)$.
\\
Here $p$ is arbitrary constant.

Any other system (\ref{a1}) with $d_1\neq0$ admitting an extension of the  principal algebra (\ref{a2})
 is reduced by an ET from (\ref{a3}) to one of the listed in cases 1)--5).
\end{theo}

\begin{remark}
Using the simple ET from (\ref{a3}), one can set $k_0k(t)$ ($k_0$ is an arbitrary constant) instead of $k(t)$ in each case of Theorem \ref{th2} without any changes in Lie symmetry operators. It is useful from the applicability point of view.
\end{remark}

\begin{theo} \label{th3}
System (\ref{a1}) with $d_1=0$ (then automatically $d_2\neq0$) admits the extension of the  principal algebra (\ref{a2}) only in four cases. In each case the additional operators have the structure
\begin{equation}\label{a13}\begin{array}{l} X=\xi^0(t,u)\,\partial_t+\xi^1(x,y)\,\partial_x+\xi^2(x,y)\,\partial_y+
\eta^1(t,u)\,\partial_u+\left(G(t,x,y,u)+(\xi^0_t-2\xi^1_x+\eta^1_u)v\right)\partial_v,\end{array}\end{equation}
where the functions $\xi^1, \ \xi^2$  form an arbitrary solution of the famous  Cauchy-Riemann system
\begin{equation}\label{a7}\medskip \xi^1_x=\xi^2_y, \ \xi^1_y=-\xi^2_x, \end{equation}
and $G$  is  an arbitrary solution of the linear first-order PDE
\begin{equation}\label{a7*}
G_t+u(a- bu^\gamma)G_u=k(t)\left(\eta^1+2u\xi^1_x-
u\eta^1_{u}\right)+uk'(t)\xi^0+u^2k(t)\left(a- bu^\gamma\right)\xi^0_u.
\end{equation}

 In the operator $X$,   the functions $\xi^0$ and $\eta^1$  depending on the parameters $\gamma$ and $a$  have the forms\,:

1) if  $\gamma\neq1$ and $a\neq0$  then  \[\xi^0=\alpha_1+\frac{\alpha_2e^{at}}{au}\left(a-a\gamma+b\gamma u^\gamma\right)\left(a-bu^\gamma\right)^{-1+\frac{1}{\gamma}}, \ \eta^1=\alpha_2e^{at}\left(a-bu^\gamma\right)^{\frac{1}{\gamma}};\]

2) if $\gamma\neq1$ and $a=0$ then  \[\xi^0=\alpha_1-\gamma\alpha_2t, \ \eta^1=\alpha_2u;\]

3) if  $\gamma=1$ and $a\neq0$ then
\begin{eqnarray} \nonumber  && \xi^0=e^{2at}\int\frac{f_1\left(\frac{a-bu}{u}\,e^{at}\right)}{u^4}\,du+g(t),
\medskip \\ \nonumber &&
\eta^1=\frac{u}{ae^{at}}\Big(f_2(\omega)+\omega^2\int e^{-3at}\left(\omega+be^{2at}\right)f_1\left(\omega e^{-at}\right)dt\Big),\ \omega=\frac{a-bu}{u}\,e^{2at};
\end{eqnarray}

4) if  $\gamma=1$ and $a=0$ then
\begin{eqnarray} \nonumber  && \xi^0=\int\frac{f_1\left(\frac{1}{u}-bt\right)}{u^4}\,du+g(t), \medskip \\ \nonumber &&
\eta^1=u\left(f_2(\omega)+b^2\int \left(\omega+2bt\right)f_1\left(\omega+bt\right)dt\right),\ \omega=\frac{1}{u}-2bt.
\end{eqnarray}
Here $\alpha_1$ and $\alpha_2$ are arbitrary parameters while the functions $f_1, \ f_2$  and $g$ are such
that the identity $\xi^0_t=-\eta^1_{u}$ should take place.

\end{theo}

Proofs of Theorems~\ref{th2}  and \ref{th3} are based on the technique, which is a combination of the classical Lie method and the group of ETs. This technique is often called the Lie--Ovsiannikov method because L.V. Ovsyannikov was the first who applied such technique for solving the LSC problem (group classification problem) for a class of the nonlinear heat equations \cite{ovs}.  Here we present the proof of Theorems~\ref{th3}  because that is more complicated comparing with the proof of  Theorem~\ref{th2}.

\emph{Proof of Theorem \ref{th3}.}

System  (\ref{a1}) with $d_1=0, \ d_2\neq0$ can be rewritten as
\begin{equation}\label{a8}\begin{array}{l} \medskip  u_t = u(a- bu^\gamma),  \\
v_t = \Delta u +k(t)u.\end{array}\end{equation}
 using an appropriate ET from (\ref{a3}).
 As usually, we start from the most  general form of a  Lie symmetry operator  \begin{equation}\label{a9} \begin{array}{l} X=\xi^0(t,x,y,u,v)\,\partial_t+\xi^1(t,x,y,u,v)\,\partial_x+\xi^2(t,x,y,u,v)\,\partial_y\\ \hskip1.5cm +
\eta^1(t,x,y,u,v)\,\partial_u+\eta^2(t,x,y,u,v)\partial_v.\end{array}\end{equation}

 In order to find all Lie symmetry operators of the form (\ref{a9}) of system (\ref{a8})  one should apply the following invariance criterion\,:
\begin{equation}\label{a10}
\begin{array}{l}
\mbox{\raisebox{-1.6ex}{$\stackrel{\displaystyle  
X}{\scriptstyle 1}$}} \left(u_t - u(a- bu^\gamma)\right)\Big\vert_{{\cal{M}}}  
 =0, \\[0.3cm]  
\mbox{\raisebox{-1.6ex}{$\stackrel{\displaystyle  
X}{\scriptstyle 2}$}} \left(v_t-\Delta u -k(t)u\right)\Big\vert_{{\cal{M}}}  
 =0,
\end{array}  \end{equation}  where  operators $ \mbox{\raisebox{-1.6ex}{$\stackrel{\displaystyle
X}{\scriptstyle 1}$}} $  and $ \mbox{\raisebox{-1.6ex}{$\stackrel{\displaystyle
X}{\scriptstyle 2}$}}$
are the first and second  prolongations of the operator $X$ and the manifold ${\cal{M}}$ is defined by the system of equations
\begin{eqnarray} \nonumber  &&  \medskip  u_t = u(a- bu^\gamma),  \ v_t = \Delta u +k(t)u, \\ && \nonumber
u_{tt} = \left(a- b(\gamma+1)u^\gamma\right)u_t, \    u_{tx} = \left(a- b(\gamma+1)u^\gamma\right)u_x,\ u_{ty} = \left(a- b(\gamma+1)u^\gamma\right)u_y.
\end{eqnarray}
It should be stressed, that the manifold ${\cal{M}}$ involves not only the equations of the system in question but also the first-order consequences of the first equation of (\ref{a8}). These consequences guarantee a  {\it complete} solving the LSC problem. Notably, such peculiarity does not occur for scalar PDEs but one was noted for some systems of PDEs involving equations of different order (see, e.g., the relevant discussion in \cite{bl-anco-10} Section 1.2.5).

Having the correctly-defined manifold ${\cal{M}}$,  the invariance criterion (\ref{a10}) after   rather
standard calculations  leads to  the system of determining equations as follows
\begin{eqnarray} \label{a14}  && \xi^0_{x}=\xi^0_{y}=\xi^0_{v}=0, \ \xi^1_{t}=\xi^1_{u}=\xi^1_{v}=0, \ \xi^2_{t}=\xi^2_{u}=\xi^2_{v}=0, \\ \label{a15} && \eta^1_x=\eta^1_y=\eta^1_v=0, \ \eta^2_v=\xi^0_t-2\xi^1_x+\eta^1_u, \\ \label{a16} &&
 \xi^1_x=\xi^2_y, \ \xi^1_y=-\xi^2_x, \\ \label{a17}  &&
\eta^1_{uu}=2\left(a-b\left(1+\gamma\right)u^\gamma\right)\xi^0_u+u\left(a- bu^\gamma\right)\xi^0_{uu},\\ \label{a18}  &&
\eta^1_{t}=\left(a-b\left(1+\gamma\right)u^\gamma\right)\,\eta^1+u^2\left(a- bu^\gamma\right)^2\xi^0_u+u\left(a- bu^\gamma\right)\left(\xi^0_t-\eta^1_{u}\right),\\  && \eta^2_t+u\left(a- bu^\gamma\right)\eta^2_u=k(t)\left(\eta^1+2u\xi^1_x-
u\eta^1_{u}\right)+uk'(t)\xi^0+u^2k(t)\left(a- bu^\gamma\right)\xi^0_u. \label{a19}
\end{eqnarray}

Equations (\ref{a14})--(\ref{a15}) can be  easily integrated, hence    the general form of the infinitesimal  operator
(\ref{a9}) can be specified as
(\ref{a13}). Now substituting the function \[\eta^2=G(t,x,y,u)+\left(\xi^0_t-2\xi^1_x+\eta^1_u\right)v\] into equation (\ref{a19}) and splitting the
equation obtained w.r.t. the variable $v$, we arrive at the linear  equation  (\ref{a7*}) for $G(t,x,y,u)$  and the equation
\begin{equation}\label{a20}u\left(a- bu^\gamma\right)\xi^0_{tu}+\xi^0_{tt}+u\left(a- bu^\gamma\right)\eta^1_{uu}+\eta^1_{tu}=0.\end{equation}
Obviously, equations (\ref{a16}) coincide with (\ref{a7}).

Thus, we need only to solve the overdetermined system of equations (\ref{a17}), (\ref{a18}) and (\ref{a20}) w.r.t. the functions $\xi^0$ and $\eta^1$. This is a nontrivial task because the function $\xi^0$ depends on the dependent variable $u$ in contrast to the standard situation for the systems of reaction-diffusion equations (see \cite{ch-king-2006,ch-da-mu-2017} and the papers cited therein). Since $\xi^0_u\not=0$,
we used the  differential consequence of equations  (\ref{a17}) and (\ref{a18}) w.r.t. the variables  $t$  and $u$. Differentiating equation (\ref{a17}) w.r.t. $t$, taking the second-order consequence  of equation (\ref{a18})  w.r.t. $u$ and making the relevant calculations, we were able to derive the simple relation
\begin{equation}\label{a21} \xi^0_{t}=\frac{\left(1-\gamma\right)\eta^1-u\eta^1_u}{u}.
\end{equation}
Substituting the derivatives $\xi^0_{tu}$ and $\xi^0_{tt}$ derived from  (\ref{a21}) into  equation (\ref{a20}) one  arrives at the classification equation
\[(1-\gamma)\Big(u(a-bu^\gamma)\eta^1_{u}-u^2(a- bu^\gamma)^2\xi^0_u+(a\gamma-a+bu^\gamma)\eta^1\Big)=0.
\]
Thus,  the following two cases  must be examined separetely\,: $\gamma\neq1$ and $\gamma=1.$

In the case $\gamma\neq1$,  one immediately obtains \begin{equation}\label{a22} \eta^1_{u}=u\left(a- bu^\gamma\right)\xi^0_u+\frac{\left(a-a\gamma-bu^\gamma\right)\eta^1}{u\left(a-bu^\gamma\right)}.\end{equation} Differentiating equation (\ref{a22}) w.r.t. the variable $u$ and substituting the expression  obtained for  $\eta^1_{uu}$ into equation (\ref{a17}), one  arrives at the equation
\begin{equation}\label{a23} \xi^0_{u}=\frac{a\left(\gamma-1\right)\eta^1}{u^2\left(a-bu^\gamma\right)^2}.\end{equation}
Equations  (\ref{a22}) and (\ref{a23}) can be easily integrated.  The general solutions are
 \begin{equation}\label{a24}\xi^0=g(t)+\frac{f(t)}{au}\left(a-a\gamma+b\gamma u^\gamma\right)\left(a-bu^\gamma\right)^{-1+\frac{1}{\gamma}}, \ \eta^1=\left(a-bu^\gamma\right)^{\frac{1}{\gamma}}f(t),\end{equation}
if $a\neq0$ and
\begin{equation}\label{a25}\xi^0=g(t), \ \eta^1=f(t)u, \end{equation}
if $a=0$. Here $f(t)$ and $g(t)$ are arbitrary smooth functions  at the moment. In order to find the functions  $f(t)$ and $g(t)$,  one needs to substitute (\ref{a24}) and (\ref{a25}) into equation  (\ref{a18}).  As a result, cases $\emph{1)}$ and $\emph{2)}$ of Theorem~\ref{th3} were identified.

In the case $\gamma=1$, equations (\ref{a17}), (\ref{a18})  and (\ref{a21}) take the forms
\begin{equation}\label{a26}\begin{array}{l} \medskip \xi^0_t=-\eta^1_{u}, \\ \medskip
\xi^0_{tu}+u(a- bu)\xi^0_{uu}+2(a-2bu)\xi^0_u=0,\\ \medskip
\eta^1_{t}+2u(a- bu)\eta^1_{u}=(a-2bu)\,\eta^1+u^2(a-bu)^2\xi^0_u,
\end{array}\end{equation} while equation (\ref{a20}) is satisfied identically. Integrating  the last
 two equations of system (\ref{a26}), we arrive exactly at Cases $\emph{3)}$ (for $a\neq0$)  and $\emph{4)}$  ($a=0$) of Theorem~\ref{th3}.
 Notably,  we used the transformations
$u^*=\frac{a-bu}{u}\,e^{at}$ (in the case $a\neq0$) and $u^*=\frac{1}{u}-bt$ (in the case $a=0$) for solving the second equation of system (\ref{a26}).

 As one can note, the functions $\xi^0$ and $\eta^1$ involve   arbitrary functions $f_1, \ f_2$ and $g$ in cases  $\emph{3)}$ and $\emph{4)}$ of Theorem~\ref{th3}. However, they  should be specified  from the equation $\xi^0_t=-\eta^1_{u}.$  At the final stage, both functions, $\xi^0$ and $\eta^1$, should be inserted into  (\ref{a7*}). The equation obtained is an integrable first-order PDE and its general solution is easily constructed in an explicit form provided   the function $k(t)$ is given. Thus, all the coefficients of operator (\ref{a13}) are identified.


\emph{The proof is now complete. }

\medskip

 Let us present examples of highly nontrivial Lie symmetries in  cases $\emph{3)}$  and  $\emph{4)}$ (see  Theorem~\ref{th3}). Setting $f_1=1$ into expressions arising in  cases $\emph{3)}$  and  $\emph{4)}$  and using the equation $\xi^0_t=-\eta^1_{u}$, one can specify   $\xi^0$ and $\eta^1$  as follows \begin{eqnarray} \nonumber  && \xi^0=\left(\frac{b^3}{3a^2}-\frac{1}{3u^3}\right)e^{2at}+\frac{\alpha_1b}{a^2}\,e^{at}+\frac{\alpha_2}{a^2}\,e^{-at}+\alpha_0, \medskip \\ \nonumber  && \eta^1=-\frac{(a+2bu)(a-bu)^2}{3a^2u^2}\,e^{2at}+\frac{\alpha_1(a-bu)}{a}\,e^{at}+\frac{\alpha_2u}{a}\,e^{-at}, \end{eqnarray}  and
\begin{eqnarray} \nonumber  && \xi^0=\frac{b^3t^3}{3}+\alpha_2bt^2-\alpha_1t+\alpha_0-\frac{1}{3u^3}, \medskip\\  \nonumber  && \eta^1=(\alpha_1-2\alpha_2bt-b^3t^2)\,u+b^2t+\alpha_2\end{eqnarray} in cases $\emph{3)}$ and  $\emph{4)}$, respectively (here $\alpha_0,\ \alpha_1, $ and $ \alpha_2$ are arbitrary constants).

Now one should  use  the functions $\xi^0$ and $\eta^1$  for finding  the function $G$ from  equation (\ref{a7*}). In order to avoid cumbersome formulae
   we additionally set $\alpha_0=\alpha_1=\alpha_2=0$ (just for simplicity), take the particular solution $\xi^1=y, \ \xi^2=-x$ of the Cauchy-Riemann system (\ref{a7})  and fix the function $k(t)$\,: $k(t)=e^{-2at}$ (case $\emph{3)}$ ) and $k(t)=t^{-2}$  (case $\emph{4)}$). As a result, one arrives at  the  Lie symmetry operators
\begin{equation}\label{a27}\begin{array}{l}
X= \left(\frac{b^3}{3a^2}-\frac{1}{3u^3}\right)e^{2at}\partial_t+y\partial_x-x\partial_y-\frac{(a+2bu)(a-bu)^2}{3a^2u^2}\,
e^{2at}\partial_u\medskip\\
\qquad +\left(\frac{2b^2}{3a^2}\,\ln u+\frac{bu-a}{3au^2}+H\left(x,y,\frac{a-bu}{u}\,e^{at}\right)\right)\partial_v
\end{array}\end{equation} and
\begin{equation}\label{a28}\begin{array}{l} X=
\left(\frac{b^3t^3}{3}-\frac{1}{3u^3}\right)\partial_t+y\partial_x-x\partial_y+b^2t\left(1-btu\right)\partial_u\medskip\\
\qquad +\left(\frac{2b^2}{3}\,\ln (tu)+\frac{btu-1}{3t^2u^2}+H\left(x,y,bt-\frac{1}{u}\right)\right)\partial_v
\end{array}\end{equation} in cases $\emph{3)}$ and  $\emph{4)}$, respectively. Here $H$ is an arbitrary smooth function.

\section{ Exact solutions and their interpretation } \label{sec:3}

Theorems \ref{th2} and \ref{th3}  allows us to reduce the basic system (\ref{a1}) to that of lower dimensionality. In fact, using the Lie symmetry operators (or their linear combinations) listed in Theorems~\ref{th2} and \ref{th3} one can  reduce (\ref{a1}) to the corresponding $(1+1)$-dimensional system and the latter to an ODE system.  Here we examine only two cases in order to show that those lead to useful exact solutions.

First of all, one may simplify the nonlinear system (\ref{a1})  using the
 ETs (\ref{a3}) with the correctly-specified parameters
\begin{eqnarray} \nonumber  &&  \medskip  \overline{t} =at, \
\overline{x}=\sqrt{\frac{a}{d_1}}\, x, \  \overline{y}=\sqrt{\frac{a}{d_1}}\, y,\\ \nonumber  &&
\overline{u}=\left(\frac{b}{a}\right)^{1/\gamma}u, \  \overline{v}=\left(\frac{b}{a}\right)^{1/\gamma}v,
\end{eqnarray}
to the form
\begin{equation}\label{3-1*}\begin{array}{l} \medskip  u_t = \Delta u+u(1- u^\gamma),  \\
v_t = D \Delta u +\frac{1}{a}\,k\left(\frac{t}{a}\right)u, \ D=\frac{d_2}{d_1}. \end{array}\end{equation}
Here and in what follows we preserve the old notations for all the variables.

\textbf{Example 1}.  Let us apply the operator $\partial_y$ from the principal algebra (\ref{a2}) for reduction of the basic system (\ref{3-1*}). Obviously, this operator produces the trivial ansatz $u=u(t,x), \  v=v(t,x)$, so that we arrive at the system
\begin{equation}\label{3-3}\begin{array}{l} \medskip  u_t = u_{xx}+u(1- u^\gamma),  \\
v_t = D u_{xx} +\frac{1}{a}\,k\left(\frac{t}{a}\right)u, \ D=\frac{d_2}{d_1}. \end{array}\end{equation}

This system is nothing else but the initial system under assumption that the distribution of the infected persons is one-dimensional in space (i.e., the diffusion w.r.t. the axis $y$ is very small). In this case, the distribution of the total  number of  deaths will be also  one-dimensional.

Making the further plausible assumption $d_1 \gg d_2$, i.e., the space diffusion of the infected persons leads mostly to increasing the total number of the COVID-19 cases and not so much to new deaths, we may put $D=0$. Following our previous paper \cite{ch-dav-2020}, we specify the function $k(t)=k_0 e^{-\alpha t}$ (hereafter $k_0>0, \  \alpha>0$). Thus, system (\ref{3-3}) takes the form
\begin{equation}\label{3-4}\begin{array}{l} \medskip  u_t = u_{xx}+u\left(1- u^\gamma\right),  \\
v_t = \frac{k_0}{a}\exp\left(-\frac{\alpha t}{a}\right)u. \end{array}\end{equation}

Using Theorem~\ref{th2} (see case \emph{4)} therein), one notes that system (\ref{3-4}) admits the Lie symmetry $\partial_t-\frac{\alpha }{a}\,v\partial_v$. So, taking the linear combination of this operator and the operator $\partial_x$
\[
X = c\partial_x + \partial_t-\frac{\alpha }{a}\,v\partial_v, \ c \in \mathbb{R},
\]
we obtain the ansatz
\begin{equation}\label{3-6}
u=\phi(\omega),\ v=\exp\left(-\frac{\alpha t}{a}\right)\psi(\omega), \ \omega=x-ct.
\end{equation}
Substituting (\ref{3-6}) into (\ref{3-4}), one arrives at the ODE system
\begin{equation}\label{3-7}\begin{array}{l} \medskip  \phi''+ c\phi'+ \phi\left(1- \phi^\gamma\right)=0, \\
c\psi'+ \frac{\alpha}{a}\psi =-\frac{k_0}{a}\phi. \end{array}\end{equation}

The first equation in (\ref{3-7}) is the known second-order ODE, which arises in many applications (e.g., for study the Fisher equation  and its natural  generalizations \cite{mur2}). The general solution of this ODE can be presented only in parametric form (see, e.g., \cite{pol-za-2018}), which is not useful for further analysis. However, an exact solution in the explicit form can be constructed for the correctly-specified parameter $c=\frac{\gamma+4}{\sqrt{2(\gamma+2)}}$. To the best of our knowledge, this parameter and the relevant solution was established in \cite{abdel-82} for the first time (see more references in \cite{gi-ke-04}).
As a result, we obtain the traveling front solution of the first equation in system (\ref{3-4}):
\begin{equation}\label{3-8}
u(t,x) =\left(1+ A\exp \left(\frac{\gamma}{\sqrt{2(\gamma+2)}}\,\omega\right)\right)^{-2/\gamma}, \  \omega=\pm\, x-\frac{\gamma+4}{\sqrt{2(\gamma+2)}}\,t,  \  A>0.
\end{equation}
Notably  (\ref{3-8}) with $A<0$ is still a solution, however, one possesses a singularity. It should be also noted that the basic system (\ref{a1})  and its particular cases  derived above are invariant under the discrete transformation $x \to -x$, therefore we may put $\omega= x-\frac{\gamma+4}{\sqrt{2(\gamma+2)}}\,t$ in what follows.

Having the function $u$ in the explicit form  (\ref{3-8}), one easily derives the function $v$  from the second equation of system (\ref{3-4}):
\begin{equation}\label{3-9}
v(t,x) =\frac{k_0}{a}\int \exp\left(-\frac{\alpha t}{a}\right)\left(1+ A\exp \left(\frac{\gamma}{\sqrt{2(\gamma+2)}}\,\omega\right)\right)^{-2/\gamma}dt  + g(x),
\end{equation}
where $g(x)$  is an arbitrary smooth function.
The integral in right hand side of (\ref{3-9}) cannot be expressed in the terms of elementary functions for arbitrary parameters $a, \ \alpha$ and $\gamma$, therefore we study below a particular case.

In order to avoid cumbersome formulae, let us set $\gamma=1$ in system (\ref{3-4}). In this case, the above exact solution takes the form
\begin{equation}\label{3-10}\begin{array}{l} \medskip
u(t,x) =\left(1+ A\exp \left(\frac{1}{\sqrt{6}}\,\omega\right)\right)^{-2}, \  \omega=x-\frac{5}{\sqrt{6}}\,t,  \\
v(t,x) =\frac{k_0}{a}\int \exp\left(-\frac{\alpha t}{a}\right)\left(1+ A\exp \left(\frac{1}{\sqrt{6}}\,\omega\right)\right)^{-2}dt  + g(x).
\end{array}\end{equation}

\begin{remark} The expression for $u(t,x)$ in (\ref{3-10}) presents the well-known traveling front of the famous Fisher equation $u_t= u_{xx}+u(1-u)$, which was firstly identified in \cite{abl-zep}.
\end{remark}

The integral in right hand side of (\ref{3-10}) can be expressed in the terms of elementary functions for several values of the parameter $\frac{\alpha }{a}$. Taking $\frac{\alpha }{a}=\frac{5}{6}$, for example, we obtain

\begin{equation}\label{3-11}\begin{array}{l} \medskip
u(t,x) =\left(1+ A\exp \left(\frac{1}{\sqrt{6}}\,\omega\right)\right)^{-2}, \  \omega=x-\frac{5}{\sqrt{6}}\,t,  \\
v(t,x) =g(x) -\frac{6k_0}{5a}\exp\left(-\frac{x}{\sqrt 6}\right)
\left(A+ \exp \left(-\frac{1}{\sqrt{6}}\,\omega\right)\right)^{-1}.
\end{array}\end{equation}

Now we turn to a possible interpretation of solution  (\ref{3-11}). First of all, the functions $u$ and $v$ should be nonnegative for any $t>0$ and $x \in \mathbb{I}$ (here $\mathbb{I} \subset \mathbb{R}$) because they represent the densities. Obviously, the functions $u$ is always positive.   It is easily seen that each function $g(x)$ satisfying the inequality
\[  g(x) \geq  \frac{6k_0}{5a}\exp\left(-\frac{x}{\sqrt 6}\right)
\left(1+ A\exp \left(-\frac{x}{\sqrt{6}}\right)\right)^{-1} \]
guarantees also  nonnegativity of $v$. In particular, one may take the function
\begin{equation}\label{3-12}
 g(x) = \frac{6k_0}{5a}\exp\left(-\frac{x}{\sqrt 6}\right)
\left(1+ A\exp \left(-\frac{x}{\sqrt{6}}\right)\right)^{-1},
\end{equation}
which guarantees that the zero density of the deaths in the initial time $t=0$, i.e., $v(0,x)=0$.

\begin{figure}[t]\begin{center}
 \includegraphics[width=6.5cm]{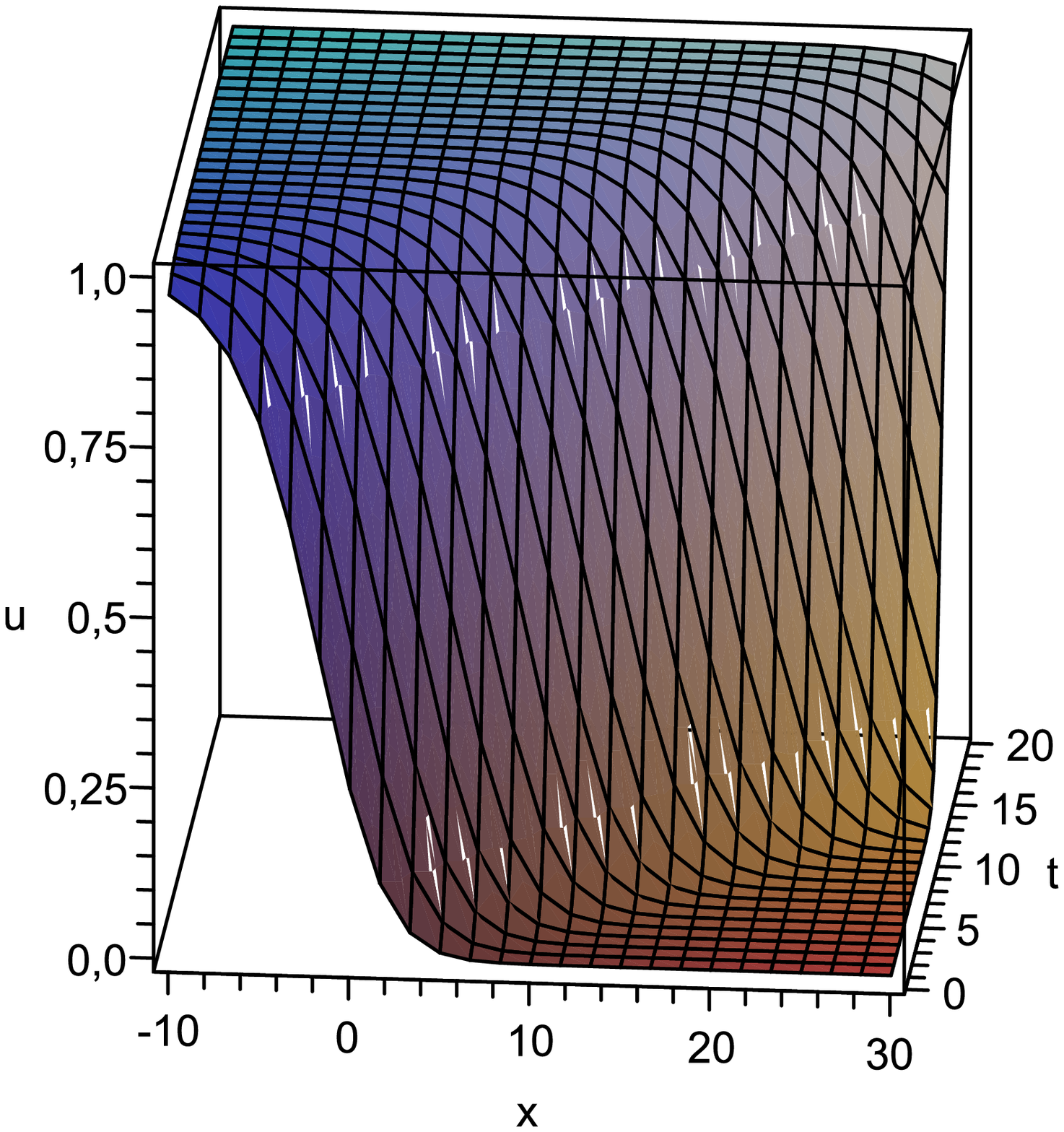}
  \includegraphics[width=6.5cm]{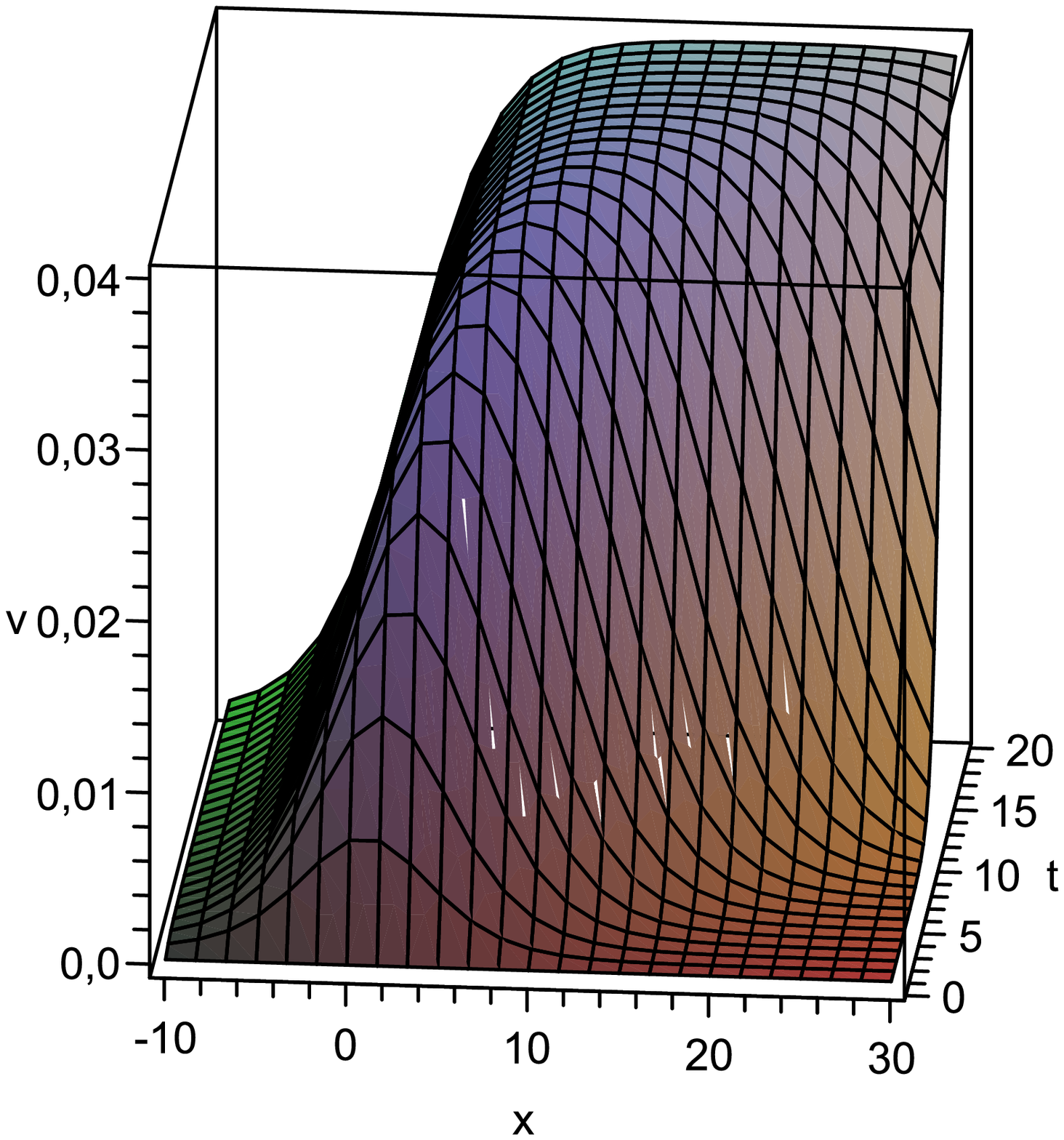}
\end{center}
 \caption{Solution  (\ref{3-11})-(\ref{3-12})  of the nonlinear  system
(\ref{3-4}) with $\gamma=1$. The function $u(t,x)$ (left surface) describes the density of the of the COVID-19 cases while the function $v(t,x)$ (right surface) describes the density of deaths. The parameters are: $k_0=0.01,  \ a=0.3, \ A=1$.
}\label{f-1}
\end{figure}
\begin{figure}[t]\begin{center}
 \includegraphics[width=6.5cm]{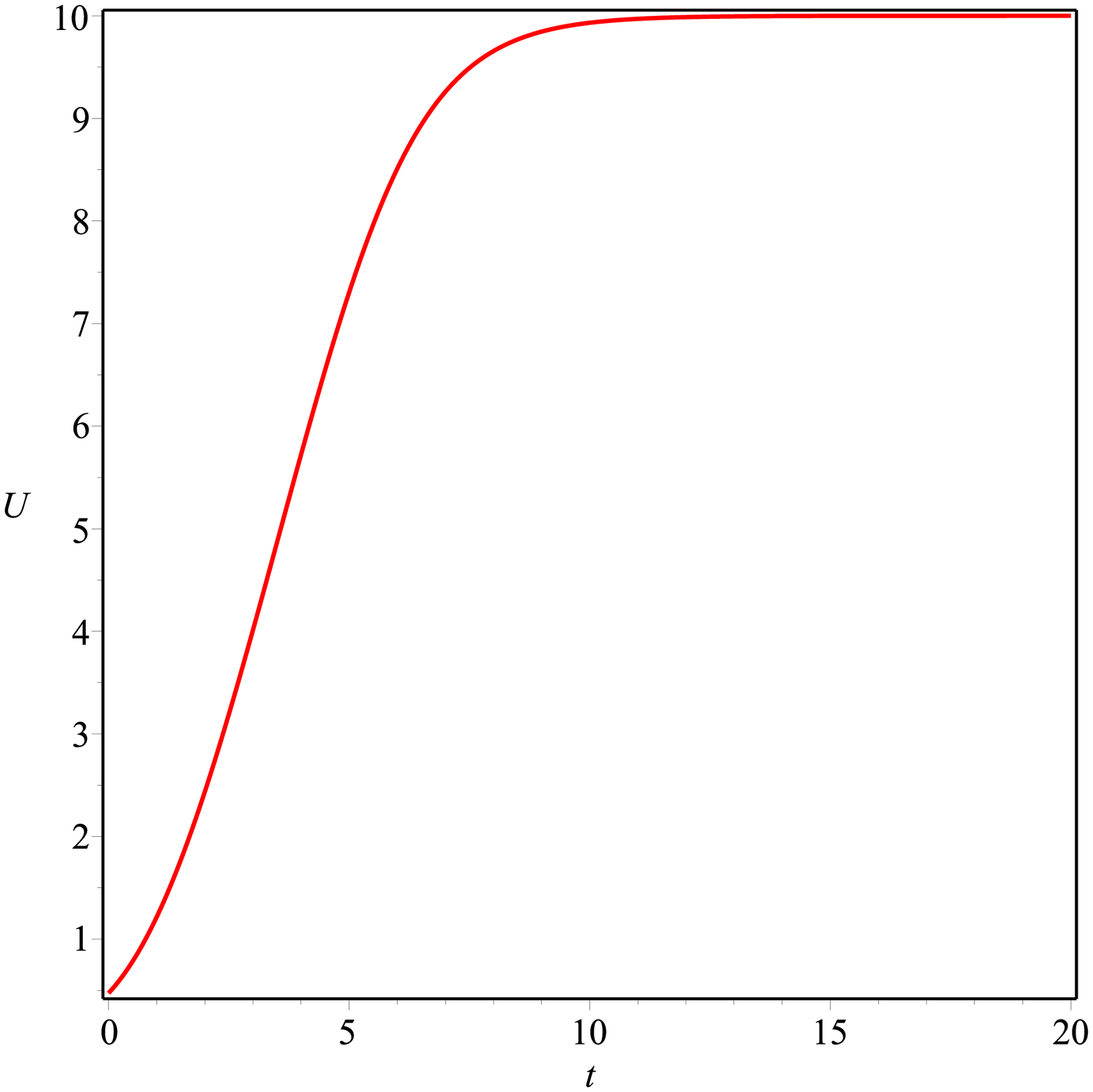}
\includegraphics[width=6.5cm]{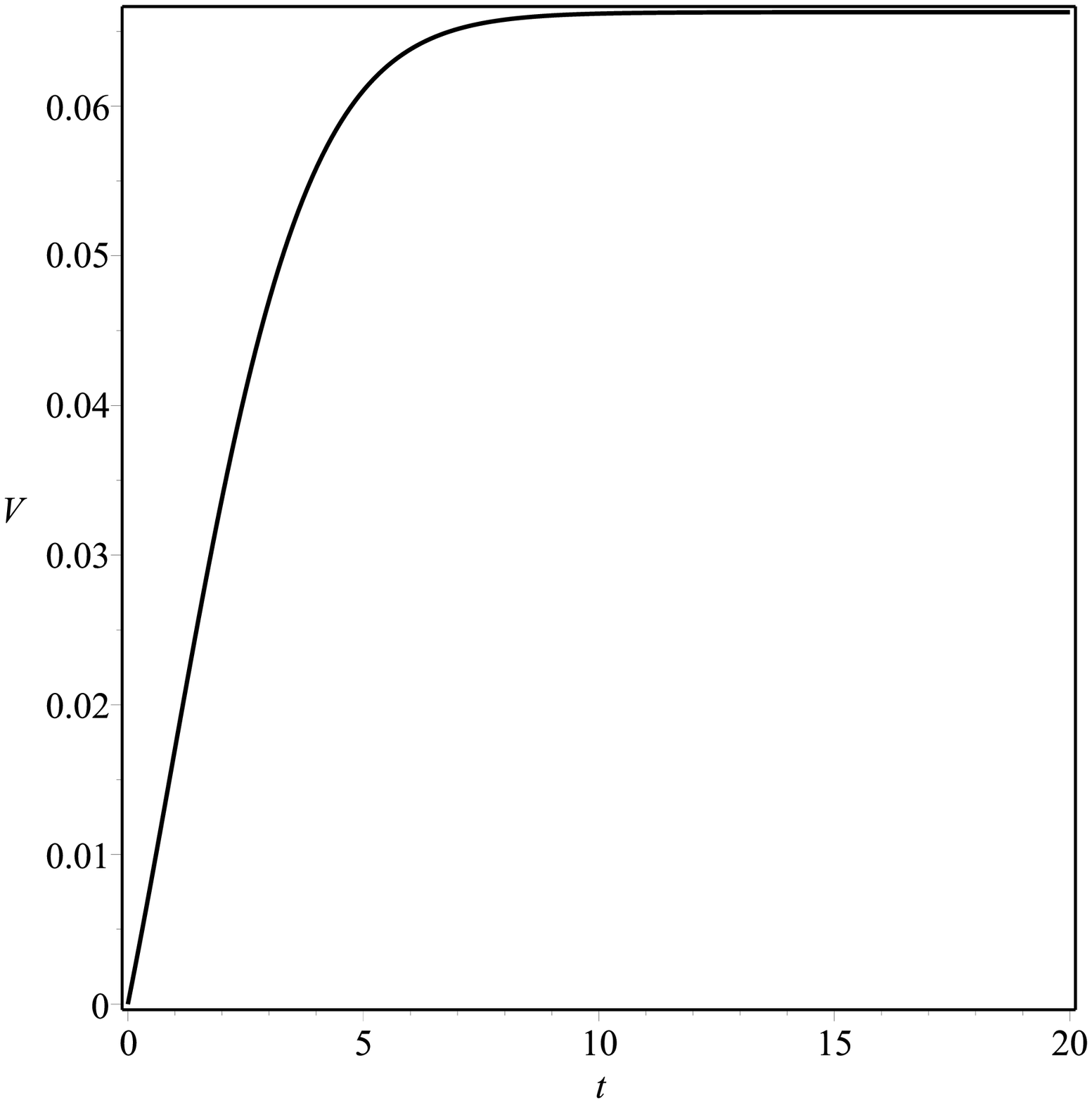}
\end{center}
 \caption{The functions  (\ref{3-14}) with  (\ref{3-12}).  The function $U(t)$ (left curve)  shows  the time-evolution of total number of the COVID-19 cases on the space interval  $[0, 10]$, while the function $V(t)$ (right curve)  shows  the time-evolution of total deaths. The parameters $k_0,  \  a$ and $ A$ are the same as in Fig.~1.
}\label{f-2}
\end{figure}

Examining the space interval $\mathbb{I}=[x_1,x_2], \ x_1<x_2$, we can calculate the total number of the COVID-19 cases and deaths on this interval as follows
\begin{equation}\label{3-13}\begin{array}{l} \medskip
U(t)=\int_{x_1}^{x_2} u(t,x)dx, \\
V(t)=\int_{x_1}^{x_2} v(t,x)dx.
\end{array}\end{equation}
So, substituting solution (\ref{3-11}) into (\ref{3-13}), we arrive at the formulae
\begin{equation}\label{3-14}\begin{array}{l} \medskip
U(t)=(x_2-x_1) - \sqrt{6}\Big[\Big(1+ A\exp \Big(\frac{x_1-\frac{5}{\sqrt{6}}\,t}{\sqrt{6}}\Big)\Big)^{-1}-\Big(1+ A\exp \Big(\frac{x_2-\frac{5}{\sqrt{6}}\,t}{\sqrt{6}}\Big)\Big)^{-1}\\
\hskip2cm +\ln\Big(1+ A\exp \Big(\frac{x_2-\frac{5}{\sqrt{6}}\,t}{\sqrt{6}}\Big)\Big)- \ln\Big(1+ A\exp \Big(\frac{x_1-\frac{5}{\sqrt{6}}\,t}{\sqrt{6}}\Big)\Big)\Big],\medskip \\
V(t)=  \int_{x_1}^{x_2} g(x)dx  - \frac{6\sqrt{6}\,k_0}{5a}e^{-\frac{5}{6}\,t}\Big[ \ln\Big(1+ A\exp \Big(\frac{\frac{5}{\sqrt{6}}\,t-x_1}{\sqrt{6}}\Big)\Big)-\ln\Big(1+ A\exp \Big(\frac{\frac{5}{\sqrt{6}}\,t-x_2}{\sqrt{6}}\Big)\Big].
\end{array}\end{equation}
Obviously the functions $U(t)$  and $V(t)$  are increasing and  bounded, because
\[ (U,V) \to \Big((x_2-x_1),  \int_{x_1}^{x_2} g(x)dx \Big) \  as \  t \to +\infty. \]
Moreover, taking
the  appropriate function $g(x)$, we  can guarantee that
\[ U(0)=U_0\geq 0,  \  V(0)=V_0\geq 0.\]
Thus, one may claim that the exact solution  (\ref{3-11}) possesses all  necessary properties for the description
of the  distribution of the COVID-19 cases  and the deaths from this virus in time-space (under 1D approximation).

Examples of  the exact solution  (\ref{3-11})  with the specified parameters  and the corresponding functions
(\ref{3-14}) are presented in Fig.~\ref{f-1} and Fig.~\ref{f-2}. Notably, the parameters $a$ and $k_0$ were taken approximately the same  as in \cite{ch-dav-2020}.
It follows from Fig.~\ref{f-1}  that the spread of the COVID-19 cases in space has the  form of a traveling  wave and this coincides (at least qualitatively) with the real situation in many countries. In Ukraine, for example, the COVID-19 pandemic started in the western part and then spread
to the central and eastern parts of  Ukraine (the major exception was only the capital Kyiv, in which the total number of  the COVID-19 cases was high from the very beginning).
 The distribution of   deaths in space has more complicated behavior (see the right plot in Fig.~\ref{f-1}). On the other hand, it is easily  seen from Fig.~\ref{f-1} and Fig.~\ref{f-2}
 that $v(t,x)\ll u(t,x)$  what is in agreement with the measured data in many countries \cite{meters}. Notably,  the behavior of the function  $v(t,x)$  can be essentially changed by the appropriate  choice of the function $g(x)$.

\medskip

\textbf{Example 2}. Let us apply the operator $y\partial_x-x\,\partial_y$ from the principal algebra (\ref{a2}) for reduction of the basic system (\ref{3-1*}). Obviously, this operator produces the well-known  ansatz $u=u(t,r), \  v=v(t,r), \ r^2=x^2+y^2$, i.e., we examine the radially-symmetric case. In this case,  we arrive at the system
\begin{eqnarray} \nonumber  && \medskip  u_t = \frac{1}{r}\left(ru_{r}\right)_r+u\left(1- u^\gamma\right),  \\ \nonumber  &&
v_t = D \frac{1}{r}\left(ru_{r}\right)_r +\frac{1}{a}\,k\left(\frac{t}{a}\right)u, \ D=\frac{d_2}{d_1}. \end{eqnarray}

Making the  same assumptions $k(t)=k_0 e^{-\alpha t}$ and  $d_1 \gg d_2$, i.e., $D=0$, as in Example~1, we obtain the system
\begin{equation}\label{3-15}\begin{array}{l} \medskip  u_t = \frac{1}{r}\left(ru_{r}\right)_r+u\left(1- u^\gamma\right),  \\
v_t = \frac{k_0}{a}\exp\left(-\frac{\alpha t}{a}\right)u. \end{array}\end{equation}

We have proved that system (\ref{3-15}) again admits the Lie symmetry $\partial_t-\frac{\alpha }{a}v\partial_v$ (however, the operator of the space translation $\partial_r$ is absent in this case). So, using this symmetry, one obtains the ansatz
\begin{equation}\label{3-16}
u=\phi(r),\ v=\exp\left(-\frac{\alpha t}{a}\right)\psi(r), \ r=\sqrt{x^2+y^2}.
\end{equation}
Substituting (\ref{3-16}) into (\ref{3-15}), one arrives at the  system
\begin{eqnarray} \nonumber  && \medskip  \phi''+ \frac{1}{r}\phi'+ \phi(1- \phi^\gamma)=0,  \\ \nonumber  &&
- \frac{\alpha}{a}\psi =\frac{k_0}{a}\phi. \end{eqnarray}
In contrast to the first equation in (\ref{3-7}), the  first ODE in the above system is much more complicated. To the best of our knowledge,  exact solutions of this equation are unknown. It can be only noted that $\phi=1$ is the steady-state  solution. As a result, we arrive at the space-homogeneous  solution of the nonlinear system  (\ref{3-15})
\[  u=1,\ v= -\frac{k_0}{\alpha}\exp\left(-\frac{\alpha t}{a}\right). \]
Of course, the restriction $\alpha<0$ should take place because the function  $v$ means the density. On the other hand, it means the exponential growing of  the function  $v$  what is rather  unrealistic because one obtains the total extinction of the population in question for a finite time.

\section{ Conclusions } \label{sec:4}

The main part of this paper is devoted to the LSC of the class of reaction-diffusion system with the cross-diffusion (\ref{a1}). The system in question was suggested in work \cite{ch-dav-2020} as the natural generalization of the mathematical model for describing the COVID-19 outbreak.

Firstly, we present a statement about the  group of ETs of system (\ref{a1}) (see Theorem~\ref{th1}) in order to establish possible relations between systems that admit equivalent invariance algebras. Secondly, we find the principal algebra of system (\ref{a1}), i.e., the maximal
invariance algebra of this system with arbitrary coefficients (see Theorem~\ref{th2*}). And lastly,  we present two main theorems (Theorems~\ref{th2} and \ref{th3}) describing  reaction-diffusion systems of the form (\ref{a1}) admitting nontrivial Lie symmetry, i.e., present the LSC of system (\ref{a1}). In Section~\ref{sec:3}, we demonstrate that the Lie symmetries identified in Section~\ref{sec:2} are useful for finding exact solutions, which can describe the spread of the COVID-19 pandemic.

From the mathematical point of view, the most interesting Lie symmetry operators of system (\ref{a1})
occur when $d_1=0$ and are  presented in Theorem~\ref{th3}.
One sees that the coefficient $\xi^0$ of the infinitesimal operator $X$ (see(\ref{a13}))  depends on the variable $u$ (excepting case $\emph{2)}$. Moreover, this dependence is nonlinear.
To the best of our knowledge, it is the first example of such dependence for  systems of evolution  equations, in particular, reaction-diffusion systems.
We assume that such unusual Lie symmetry   of system (\ref{a1})
with  $d_1=0$  can be  a consequence of its integrability. In fact, one may consider the first equation as an ODE with the variables  $x$ and $y$  as parameters. Solving this ODE, one obtains
\[
 u(t,x,y)=a^{1/\gamma}e^{a t}C(x,y)\,\Big(a+b\,C^{\gamma}(x,y)\left(e^{a \gamma t}-1\right)\Big)^{-1/\gamma},\]
where $C(x,y)$ is an arbitrary function. Substituting this expression for  $u$ into the second equation of the system, one again obtains the integrable  ODE to find the function $v$.

Finally, it should be pointed out that Lie symmetries operators, which are nonlinear w.r.t. unknown functions were recently identified for a simplification of the Shigesada--Kawasaki--Teramoto system
in \cite{ch-da-mu-2017} (see Section 3). Such peculiarity of Lie symmetry also occurs for a special Schr\"odinger type equation \cite{fu-ch-cho-96}, which can be rewritten in the form of a reaction-diffusion system with the cross-diffusion.
However, the coefficient $\xi^0$ (see operator (\ref{a9}))
  in all known Lie symmetries of a wide range of reaction-diffusion systems \cite{ch-da-mu-2017, fu-ch-cho-96, niki-05,ch-wil-96,tor-tra-val-96,stew-broad-goard-02, serov-et-al-15} (see more references in Chapter 2 of \cite{ch-da-2017})  do not depend on the unknown function(s)    in contrast to those in Theorem~\ref{th3} and examples  (\ref{a27})--(\ref{a28}). Moreover, we may conclude that the well-known `people theorem'  stating that the coefficient $\xi^0$ in each  Lie symmetry of an arbitrary scalar evolution PDE of the order two and higher   can depend only  the time variable (no dependence on space variables and/or dependent variable!), cannot be generalized on the systems of evolution equations  without additional restrictions. The problem how to define these restrictions is an open question.

From the applicability point of view, the most interesting system of the form (\ref{a1}) admitting nontrivial Lie symmetry is presented in  case $\emph{4)}$ of Theorem~\ref{th2}. Here the function $k(t)$ of system (\ref{a1}) has the form that can be useful for describing the COVID-19 outbreak \cite{ch-dav-2020}. Moreover, the diffusivity $d_2=0$ as it is stated  in \cite{viguerie-2020}. In Section~\ref{sec:3}, we  demonstrate how the Lie symmetries obtained can be applied
for constructing of exact solutions. Furthermore we  prove that an exact solution (with correctly-specified parameters)  possesses all  necessary properties for the description of the  distribution of the COVID-19 cases  and the deaths from this virus in time and space. Although it was done under 1D space approximation, this solution can be useful for the prediction of the COVID-19 pandemic if its  spread   has a favorite  direction (a typical example is Ukraine). Of course, one needs to identify all the  parameters in system (\ref{a1})  in order to calculate correct numbers of the COVID-19 cases  and  make a plausible prediction  but this lays beyond the scope of this work.


\begin{thebibliography}{99}

\bibitem{luo} X. Luo et al.,  Analysis of potential risk of COVID-19 infections in China based on a pairwise epidemic model. \emph{Preprints} (2020) doi:10.20944/preprints202002.0398.v1.

\bibitem {china-19-02-20}  L. Peng et al.,  Epidemic analysis of COVID-19 in China by dynamical modeling.  \emph{ArXiv} (2020)  arXiv:2002.06563.

\bibitem{shao} N. Shao et al., Dynamic models for coronavirus disease 2019 and data analysis. \emph{Math. Meth. Appl. Sci.} \textbf{43} (2020) 4943--4949.

\bibitem{tian} J. Tian et al.,  Modeling analysis of COVID-19 based on morbidity data in Anhui, China.  \emph{MBE} \textbf{17} (2020) 2842--2852.

 \bibitem{efim-ushi}  D. Efimov, U. Ushirobira, On interval prediction of COVID-19 development based on a SEIR epidemic model.
Research report. Inria Lille Nord Europe--Laboratoire CRIStAL--Universite de: Lille, France (2020).

 \bibitem{roda-michaelLi} W.C.  Roda,  M.B. Varugheseb, D. Han, M.Y. Li,  Why is it difficult to accurately predict the COVID-19 epidemic?  \emph{Infectious Disease Modelling} \textbf{5} (2020) 271--281.

 \bibitem{ch-dav-preprint20} R. Cherniha, V. Davydovych, A mathematical model for the COVID-19   outbreak.  \emph{ArXiv} (2020)  arXiv:2004.01487v2.

     \bibitem{meters} Available online: \emph{https://www.worldometers.info/coronavirus} (accessed on 2 December 2020).

     \bibitem {brauer-12} F. Brauer,  C. Castillo-Chavez,
\emph{Mathematical Models in Population Biology and Epidemiology}. Springer, New York, 2012.

\bibitem {k-r-2008} M.J. Keeling, P. Rohani,  \emph{Modeling Infectious Diseases in Humans
and Animals}. Princeton University Press, Princeton, 2008.

\bibitem {diekman-hees-2000}  O. Diekmann, J. Heesterbeek, \emph{Mathematical Epidemiology of Infectious Diseases:
Model Building, Analysis and Interpretation}. Chichester, John Wiley, 2000.


 \bibitem {mur2003} J.D. Murray,  \emph{Mathematical Biology  II: Spatial
 Models and Biomedical Applications}. Springer, Berlin, 2003.


\bibitem {kermack-1927} W.O. Kermack, A.G. McKendrick,   A contribution to the mathematical theory of epidemics. \emph{Proc. Roy. Soc. A} \textbf{115} (1927) 700--721.

\bibitem {anderson} R.M. Anderson, R.M. May,    Directly  transmitted  infectious diseases: control  by
vaccination. \emph{Science} \textbf{215} (1982) 1053--1060.

\bibitem {dietz} K. Dietz,   The  incidence  of  infectious  diseases under  the  influence  of  seasonal
fluctuations.  \emph{Lecture  Notes  in  Biomathematics}  \textbf{11}, Springer, Berlin,  1976, pp. 1--15.

\bibitem {viguerie-2020}  A. Viguerie et al., Simulating the spread of COVID-19 via a spatially-resolved
susceptible-exposed-infected-recovered-deceased (SEIRD) model with
heterogeneous diffusion. \emph{Appl. Math. Lett.} \textbf{111} (2021) 106617, 9 pp.

\bibitem {mammeri-2020} Y. Mammeri,
A reaction-diffusion system to better comprehend the
unlockdown: application of SEIR-type model with
diffusion to the spatial spread of COVID-19 in France.  \emph{Comput. Math. Biophys.} \textbf{8} (2020) 102--113.

 \bibitem{ch-dav-2020} R. Cherniha, V. Davydovych,  A mathematical model for the COVID-19 outbreak and its applications. \emph{Symmetry}  \textbf{12} (2020), 990, 12 pp.


\bibitem{ovs} L.V. Ovsiannikov, \emph{The Group Analysis of
 Differential Equations}.  Academic Press, New York, 1980.

    \bibitem {ch-se-pl-book}  R. Cherniha, M. Serov,  O. Pliukhin,
\emph{Nonlinear Reaction-Diffusion-Convection Equations: Lie and
Conditional Symmetry, Exact Solutions and their Applications}.
Chapman and Hall/CRC, New York,  2018.


\bibitem{bl-anco-10} G.W. Bluman, A.F. Cheviakov,
S.C.  Anco,   \emph{Applications of Symmetry
Methods to Partial Differential Equations}. Springer, New York, 2010.

   \bibitem{akh-gaz-ibr91} I.S. Akhatov, R.K. Gazizov, N.H. Ibragimov,  Nonlocal symmetries. Heuristic approach.
  \emph{J. Sov. Math.}  \textbf{55} (1991) 1401--1450.

\bibitem{ibr-tor-val91} N.H. Ibragimov, M. Torrisi, A. Valenti, Preliminary group classification of equations $v_{tt} = f(x, v_x) v_{xx} + g(x, v_x)$. \emph{J. Math. Phys.} \textbf{32} (1991) 2988--2995.




 \bibitem{ch-king-2006}	R. Cherniha,   J.R. King,   Lie  symmetries and conservation laws  of nonlinear multidimensional reaction-diffusion systems  with variable diffusivities. \emph{IMA J. Appl. Math.} \textbf{71} (2006) 391--408.

\bibitem{ch-da-mu-2017} R. Cherniha,  V. Davydovych, L. Muzyka, Lie symmetries of the
Shigesada--Kawasaki--Teramoto system. \emph{Commun. Nonlinear Sci. Numer. Simulat.} \textbf{45} (2017)  81--92.

\bibitem {mur2}  J.D. Murray,  \emph{Mathematical Biology}.  Springer, Berlin, 1989.

\bibitem{pol-za-2018}  A.D. Polyanin, V.F. Zaitsev,
\emph{Handbook of Ordinary Differential Equations for Scientists and Engineers}.  CRC Press Company, Boca Raton, 2018.

\bibitem{abdel-82} M.A. Abdelkader, Travelling wave solutions for a generalized Fisher equation. \emph{J. Math. Anal. Appl.} \textbf{85} (1982) 287--290.

\bibitem{gi-ke-04} B.  Gilding, R. Kersner,
\emph{Travelling Waves in Nonlinear Diffusion-Convection Reaction}. Birkh\"auser, Basel,
2004.

\bibitem{abl-zep} M. Ablowitz, A.  Zeppetella, Explicit solutions of Fisher's
equation for a special wave speed. \emph{Bull. Math. Biol.} \textbf{41} (1979)
835--840.

\bibitem{fu-ch-cho-96} W. Fushchych, R. Cherniha,  V. Chopyk, On unique
symmetry of two nonlinear generalizations of the Schr\"odinger equation. \emph{J. Nonlinear Math. Phys}. \textbf{3} (1996) 296--301.

\bibitem {niki-05} A.G. Nikitin,  Group classification of systems of non-linear reaction-diffusion
 equations. \emph{Ukrainian Math. Bull.} \textbf{2} (2005) 153--204.

\bibitem{ch-wil-96} R. Cherniha, H. Wilhelmsson,  Symmetry and exact solution of  heat-mass transfer equations in thermonuclear plasma.  \emph{Ukrainian Math. J.}   \textbf{48} (1996) 1434--1449.

  \bibitem {tor-tra-val-96} M. Torrisi, R. Tracina, A. Valenti,  A group analysis approach for a nonlinear differential
system arising in diffusion phenomena. \emph{J. Math. Phys.} \textbf{37} (1996) 4758--4767.

\bibitem {stew-broad-goard-02} J.M. Stewart, P. Broadbridge, J.M. Goard, Symmetry analysis and numerical modelling of invasion by malignant tumour tissue. \emph{Nonlinear Dynamics} \textbf{28} (2002) 175--193.


   \bibitem{serov-et-al-15} M.I. Serov, T.O. Karpaliuk, O.G. Pliukhin, I.V. Rassokha,  Systems of reaction-convection-diffusion equations invariant under Galilean algebras. \emph{J. Math. Anal. Appl.} \textbf{422} (2015)  185--211.


       \bibitem {ch-da-2017} R. Cherniha, V. Davydovych,
Nonlinear Reaction-Diffusion Systems --- Conditional Symmetry, Exact
Solutions and their Applications in Biology,  \emph{Lecture Notes in
Mathematics} \textbf{2196}, Springer, Cham,  2017.

\end{thebibliography}
\end{document}